\documentclass[preprint,prl,showpacs,floatfix]{revtex4}

\usepackage{graphicx}
\usepackage{amsmath}

\begin{document}

\title
{
Rate equation model of phototransduction into the membranous disks of mouse rod cells
}

\author
{
Rei Takamoto$^1$, Hiraku Nishimori$^{1,2}$,  Akinori Awazu$^{1,2}$
}

\affiliation
{
$^1$Department of Mathematical and Life Sciences, Hiroshima University,
$^2$Research Center for Mathematics on Chromatin Live Dynamics. \\
Kagami-yama 1-3-1, Higashi-Hiroshima 739-8526, Japan.
}

\date{\today}

\begin{abstract}
A theoretical model was developed to investigate the rod phototransduction process in the mouse. In particular, we explored the biochemical reactions of several chemical components that contribute to the signaling process into/around the membranous disks in the outer segments of the rod cells. We constructed a rate equation model incorporating the molecular crowding effects of rhodopsin according to experimental results, which may hinder the diffusion of molecules on the disk membrane. The present model could effectively reproduce and explain the mechanisms of the following phenomena observed in experiments. First, the activations and relaxation of the wild-type mouse rod cell progressed more slowly than those of mutant cells containing half the amount of rhodopsin on the disk membrane. Second, the strong photoactivated state of the cell was sustained for a longer period when the light stimuli were strong. Finally, the lifetime of photoactivation exhibited a logarithmic increase with increasing light strength given exposure to strong light stimuli.
\end{abstract}


\maketitle

\section{Introduction}
Most living systems have evolved the ability to sense and adapt to environmental fluctuations through several internal biochemical processes. Photoreception is one of the most important signaling processes for several higher organisms. This signaling process starts from activation of the photoreceptor rhodopsin on the membranous disk, which is regulated by negative feedback through activation and repression cascades mediated by several membrane-associated proteins, low-molecular weight signaling molecules, and ions in each rod cell\cite{rod1,rod2,rod3,rod4}.

 Recent studies have demonstrated the phenomenon of the crowding of macromolecules in individual cells, in which the volume fraction of macromolecules is much higher than that observed under typical in vitro conditions \cite{crowd1,crowd2,crowd3,crowd4,crowd5,crowd6,crowd7,crowd8,crowd9,crowd10,crowd11,crowd12,crowd13,crowd14,crowd15,crowd16,crowd17,crowd18,crowd19,crowd20}. Such ``molecular crowding'' is considered to highly suppress the diffusion of molecules\cite{crowd2,crowd3,crowd4,crowd5,crowd6,crowd7,crowd8,crowd9,crowd10,crowd11,crowd12,crowd13,crowd14} ,which ultimately hinders intra-cellular reaction processes but also often enhances protein folding \cite{crowd15,crowd16}, stabilization of the intracellular architecture \cite{crowd17,crowd18}, and processive phosphorylations of ERK mitogen-associated protein kinase and associated gene transcription\cite{crowd19,crowd20,mem1}. In a molecular crowding scenario, the volume fractions of macromolecules on the membranes of cells and in intra-cellular organelles are also considered to be as high as that in the cytoplasm \cite{mem1,mem2,mem3,mem4,mem5,mem6}. Thus, molecular crowding on the cell membrane is expected to have more complex effects for tightly controlled biochemical processes such as the pattern formations of signaling proteins \cite{mem6}.

A high volume fraction of the photoreceptor rhodopsin tends to be maintained in the outer segments of the membranous disks of the rod cells of vertebrates\cite{rod1,rod2,rod3,rod4,rod5,rod6,rod7,rod8}. Indeed, this high packing density of photoreceptors has been proposed to ensure efficient photon capture. On the other hand, recent experimental studies of mouse rod cells showed that activation and relaxation of this signaling pathway in wild type (WT) mouse rod cells was slower than those of mutant cell containing half the amount of rhodopsin on the disk membrane. This fact suggests that the crowding of rhodopsin might actually limit the diffusion of signaling proteins on the disk membrane to slow down the phototransduction processes after rhodopsin activation. \cite{rod4,rod7}.

Recently, theoretical and experimental studies revealed the influence of the crowding of rhodopsin or other molecules on the disk membrane on the activation of the phototransduction process\cite{rod6,rod8,rod9}. Moreover, the influence of the secondary messengers in the cytoplasm, such as $cGMP$, disk membrane shape, and multi-level phosphorylations of rhodopsin on the phototransductions have also been explored with reaction diffusion models\cite{rod7,rod10,rod11,rod12,rod13,rod14}. These studies have successfully depicted the photoresponse behaviors in response to weak light stimuli. However, some of the mechanisms to explain characteristic phenomena observed in experiments have not yet been sufficiently explained yet, including: (i) activation and relaxation of the WT mouse rod cell progresses more slowly than those of mutant cells containing half the amount of rhodopsin on the disk membrane; (ii) the time interval of the photoactive state is drastically prolonged in response to strong light stimuli compared to weak stimuli; (iii) the lifetime of photoactivation exhibits a logarithmic increase with light strength under strong light stimuli.

To address these questions, in this study, we constructed a theoretical model of the whole process of phototransduction occurring in the outer segment of WT and mutant mouse rod cells in order to unveil the characteristics and the mechanism underlying the observed photoresponse behaviors for a wide range of light intensities. Recent experiments have shown that $\sim 20$ types of signaling components such as proteins, low-molecular weight molecules, ions, and their complexes contribute to these processes. Based on these experimental findings, we constructed a rate equation model incorporating the molecular crowding effects of rhodopsin that could hinder the diffusion and reactions of molecules on the disk membrane. We confirmed that the present model can qualitatively reproduce and explain the light intensity-dependent temporal phototransduction behaviors, lifespan of activation, and rhodopsin volume fraction-dependent restrictions of the phototransduction speed in mouse rod cells, as observed in experiments.

\section{Model and methods}
\subsection{Signaling components and reaction network of phototransduction}
The molecular networks of vertebrate phototransduction have been extensively investigated experimentally \cite{rod1,rod2,rod3,rod4}. After activation of rhodopsin ($R$) as $R \to R^*$ ($^*$ indicates the active form) by a light stimulus, the signal goes through a series of activation and relaxation states via the following reactions processes (Fig. 1).

\begin{figure}
\begin{center}
\includegraphics[width=12.0cm]{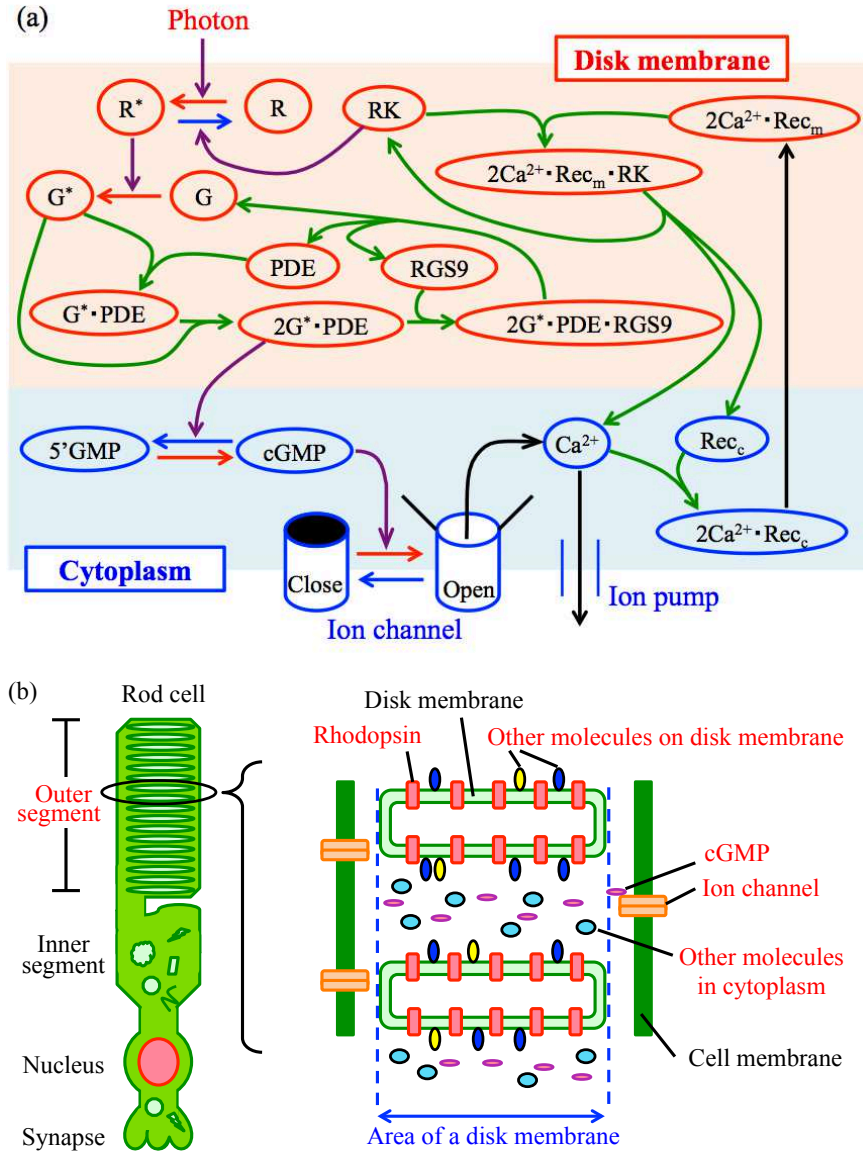}
\end{center}
\caption{{\bf Illustration of the phototransduction network in the outer segment of a mouse rod cell: } (a) Each circle represents the chemical components of the signaling transduction processes, where red and blue circles indicate chemical components on and outside of the disk membrane, respectively. Arrows represent the reaction process, where red (blue) arrows indicate the transition from the inactive (active) to active (inactive) state of chemical components, such as activation (inactivation), synthesis (degradation), and opening (closing); green arrows indicate the binding and dissociation of chemical components; purple arrows indicate the catalyzations; and black arrows indicate the movement of chemical components. (b) Illustrations of the local part of the outer segment and representation of the relative concentrations of each type of signaling component.}
\end{figure}

Phototransduction starts with the activation of the G-protein transducin ($G$), which diffuses on the disk membrane via collision with photoactivated rhodopsin ($R^*$), as
\begin{eqnarray}
R^* + G \to R^* + G^*,
\end{eqnarray}
where $R^*$ catalyzes the exchange of a GTP for a GDP on $G$.
$G^*$ then diffuses along the membrane surface, and two $G^*$ molecules eventually bind to a single cyclic GMP phosphodiesterase (PDE) to construct a complex, $(2G^* \cdot PDE)$, as
\begin{eqnarray}
2G^* + PDE \rightleftharpoons G^* + (G^* \cdot PDE) \to (2G^* \cdot PDE).
\end{eqnarray}
A membrane-associated regulator of G-protein signaling, $RGS9$, binds to $(2G^* \cdot PDE)$ as
\begin{eqnarray}
(2G^* \cdot PDE) + RGS9 \to (2G^* \cdot PDE \cdot RGS9).
\end{eqnarray}
Then, the G-bound GTP molecules are hydrolyzed and this complex is dissociated as
\begin{eqnarray}
(2G^* \cdot PDE \cdot RGS9) \to 2G +  PDE + RGS9.
\end{eqnarray}

$cGMP$ is synthesized by guanylate-cyclase-activating proteins in the cytoplasm
\begin{eqnarray}
 \to cGMP,
\end{eqnarray}
which opens the $Ca^{2+}$ channel ($Ch$) as
\begin{eqnarray}
n\,cGMP + Ch_c \to n\,cGMP + Ch_o, \,\,\,\,\,\,\,\, Ch_o \to Ch_c.
\end{eqnarray}
Here, $Ch_c$ and $Ch_o$ indicate the closed and open channel, respectively, and $n$ refers to the number of $cGMP$ molecules required to open a channel\cite{rod1,rod2,rod3}. In this study, we assume $n=2$. The hydrolysis of $cGMP$ is accelerated by $(2G^* \cdot PDE)$ as
\begin{eqnarray}
cGMP + (2G^* \cdot PDE) \to (2G^* \cdot PDE) \\
\end{eqnarray}
which results in the closure of $Ca^{2+}$ channels in the membrane.

$Ca^{2+}$ flows into the cytoplasm of the outer segment through the $Ca^{2+}$ channel and is extruded by an $Na^+/Ca^{2+},K^+$ exchanger as
\begin{eqnarray}
Ch_o \to Ch_o + Ca^{2+}, \,\,\,\,\,\,\,\, Ca^{2+} \to .
\end{eqnarray}
If the concentrations of $Ca^{2+}$ become reduced in the outer segment, the rod cell moves into a hyperpolarization state, which can activate the downstream neurons.

Two $Ca^{2+}$s bind to the cytoplasmic protein recoverin ($Rec_c$) as
\begin{eqnarray}
2Ca^{2+} + Rec_c \to Ca^{2+} + (Ca^{2+} \cdot Rec_c) \to 2Ca^{2+} \cdot Rec_c
\end{eqnarray}
and bind to the membrane as
\begin{eqnarray}
 2Ca^{2+} \cdot Rec_c \to 2Ca^{2+} \cdot Rec_m
\end{eqnarray}
($Rec_m$ indicates $Rec$ on membrane.).

Rhodopsin kinase ($RK$) accelerates the inactivation of $R$ through inducing the phospholyration of activated rhodopsin $R^*$ as
\begin{eqnarray}
RK + R^* \to RK + R,
\end{eqnarray}
which triggers the relaxation of this phototransduction. $RK$ is then inactivated by coupling with $(2Ca^{2+} \cdot Rec_m)$ as
\begin{eqnarray}
RK + 2Ca^{2+} \cdot Rec_m \to (2Ca^{2+} \cdot Rec_m \cdot RK),
\end{eqnarray}
and eventually returns to the active form through dissociation of the complex as
\begin{eqnarray}
(2Ca^{2+} \cdot Rec_m \cdot RK) \to 2Ca^{2+} + Rec_c + RK.
\end{eqnarray}

\subsection{Rate equation model}
Recent experiments have mainly focused on the activation and relaxation of cells in response to an instantaneous light stimulus under dark conditions \cite{rod1,rod2,rod3, rod4,rod7}. Thus, we here simulate the same general situations using the following rate equation model.

The density of signaling components $i$ ($[i]$) on the surface of a disk membrane in the outer segment of a rod cell is given as i) $[i]=$ [Number of component $i$ on one surface of the disk membrane] / [Area of the disk membrane] for signaling components on the disk membrane, and ii) $[i]=$ [Half of the number of component $i$ between two neighboring membrane disks] / [Area of the disk membrane] for signaling components in the cytoplasm (Fig. 1(b)). Here, the density of the $Ca^{2+}$ channel is given by [Number of $Ca^{2+}$ channels] / [Area of the cell membrane]. It is noted that each signaling component is assumed to occupy a finite area on the disk membrane. Thus, the total concentration of signaling components on the disk membrane is limited.

Now, for simplicity, we define the concentration of signaling component $i$, $X_i$, as a rescaled  density of signaling component $i$ obeying the limitations
\begin{eqnarray}
\rho = X^t_{R} + X^t_{G} + X^t_{PDE} + X^t_{RGS9} + X^t_{RK} 
+ X_{2Ca^{2+}\cdot Rec^m} + X_{2Ca^{2+}\cdot Rec^m\cdot RK} \le 1.
\end{eqnarray}
where $X^t_{i}$ indicates the total concentration of signaling components $i$ (rhodopsin, G-protein, $PDE$, $RGS9$, $RK$, and $Rec$) on the disk membrane.

Based on the reaction processes described in the previous subsection, the rate equations of $X_i$ after instantaneous exposure to a light stimulus are obtained as
\begin{eqnarray}
{\dot X_{R^*}} = -AX_{R^*}X_{RK},
\end{eqnarray}
\begin{eqnarray}
{\dot X_{R}} = AX_{R^*}X_{RK},
\end{eqnarray}
\begin{eqnarray}
{\dot X_{G^*}} = BX_{R^*}X_{G}-C^{+}X_{PDE}X_{G^*}   
+C^{-}X_{G^*\cdot PDE}-DX_{G^*}X_{G^*\cdot PDE},
\end{eqnarray}
\begin{eqnarray}
{\dot X_{G^*\cdot PDE}} = C^{+}X_{PDE}X_{G^*}-C^{-}X_{G^*\cdot PDE}  
-DX_{G^*}X_{G^*\cdot PDE},
\end{eqnarray}
\begin{eqnarray}
{\dot X_{2G^*\cdot PDE}} = DX_{G^*}X_{G^*\cdot PDE} 
-EX_{RGS9*}X_{2G^*\cdot PDE},
\end{eqnarray}
\begin{eqnarray}
{\dot X_{2G^*\cdot PDE \cdot RGS9}} = EX_{RGS9*}X_{2G^*\cdot PDE} 
-FX_{2G^*\cdot PDE \cdot RGS9},
\end{eqnarray}
\begin{eqnarray}
{\dot X_{G}} = 2FX_{2G^*\cdot PDE \cdot RGS9}-BX_{R^*}X_{G},
\end{eqnarray}
\begin{eqnarray}
{\dot X_{PDE}} = FX_{2G^*\cdot PDE \cdot RGS9}-C^{+}X_{PDE}X_{G^*}  
+C^{-}X_{G^*\cdot PDE},
\end{eqnarray}
\begin{eqnarray}
{\dot X_{RGS9}} = FX_{2G^*\cdot PDE \cdot RGS9}-EX_{RGS9*}X_{2G^*\cdot PDE},
\end{eqnarray}
\begin{eqnarray}
{\dot X_{cGMP}} = G-HX_{2G^*\cdot PDE}X_{cGMP}-IX_{cGMP},
\end{eqnarray}
\begin{eqnarray}
{\dot X_{Ch_o}} = JX_{cGMP}X_{Ch_c} -KX_{Ch_o},
\end{eqnarray}
\begin{eqnarray}
{\dot X_{Ca^{2+}}} = MX_{Ch_o}- NX_{Ca^{2+}}-PX_{Ca^{2+}}X_{Rec^c}  \nonumber \\
-QX_{Ca^{2+}}X_{Ca^{2+}\cdot Rec^c}+2RX_{2Ca^{2+}\cdot Rec^m \cdot RK},
\end{eqnarray}
\begin{eqnarray}
{\dot X_{Ca^{2+}\cdot Rec^c}} = PX_{Ca^{2+}}X_{Rec^c}-QX_{Ca^{2+}}X_{Ca^{2+}\cdot Rec^c},
\end{eqnarray}
\begin{eqnarray}
{\dot X_{2Ca^{2+}\cdot Rec^c}} = QX_{Ca^{2+}}X_{Ca^{2+}\cdot Rec^c}-SX_{2Ca^{2+}\cdot Rec^c},
\end{eqnarray}
\begin{eqnarray}
{\dot X_{2Ca^{2+}\cdot Rec^m}} = SX_{2Ca^{2+}\cdot Rec^c}-TX_{2Ca^{2+}\cdot Rec^m}X_{RK},
\end{eqnarray}
\begin{eqnarray}
{\dot X_{Rec^c}} = RX_{2Ca^{2+}\cdot Rec^m\cdot RK} - PX_{Ca^{2+}}X_{Rec^c},
\end{eqnarray}
\begin{eqnarray}
{\dot X_{RK}} = RX_{2Ca^{2+}\cdot Rec^m\cdot RK} -TX_{2Ca^{2+}\cdot Rec^m}X_{RK},
\end{eqnarray}
and
\begin{eqnarray}
{\dot X_{2Ca^{2+}\cdot Rec^m\cdot RK}} = TX_{2Ca^{2+}\cdot Rec^m}X_{RK} 
-RX_{2Ca^{2+}\cdot Rec^m\cdot RK}
\end{eqnarray}
Here, since the time scales of gene expression and protein degradation are much slower than those of the phenomena of focus, we assume conservation of the total concentrations of the following molecules, $X^t_i$, in the outer segment of the rod cell:
\begin{eqnarray}
X^t_{R} = X_{R} + X_{R^*}
\end{eqnarray}
\begin{eqnarray}
X^t_{G} = X_{G}+X_{G^*}+X_{G^*\cdot PDE} + 2X_{2G^*\cdot PDE} 
+ 2X_{2G^*\cdot PDE\cdot RGS9},
\end{eqnarray}
\begin{eqnarray}
X^t_{PDE} = X_{PDE} + X_{G^*\cdot PDE} + X_{2G^*\cdot PDE} 
+ X_{2G^*\cdot PDE\cdot RGS9},
\end{eqnarray}
\begin{eqnarray}
X^t_{RGS9} = X_{RGS9} + X_{G^*\cdot PDE} + X_{2G^*\cdot PDE\cdot RGS9},
\end{eqnarray}
\begin{eqnarray}
X^t_{RGS9} = X_{RGS9} + X_{G^*\cdot PDE} + X_{2G^*\cdot PDE\cdot RGS9},
\end{eqnarray}
\begin{eqnarray}
X^t_{Ch} = X_{Ch_o} + X_{Ch_c}
\end{eqnarray}
\begin{eqnarray}
X^t_{Rec} = X_{Rec^c} + X_{Ca^{2+}\cdot Rec^c} + X_{2Ca^{2+}\cdot Rec^c} 
 + X_{2Ca^{2+}\cdot Rec^m} + X_{2Ca^{2+}\cdot Rec^m\cdot RK},
\end{eqnarray}
and
\begin{eqnarray}
X^t_{RK} = X_{RK} + X_{2Ca^{2+}\cdot Rec^m\cdot RK}.
\end{eqnarray}

\subsection{Implemetation of light stimuli and evaluation of photoactivation of the cell}
In the present arguments, the instantaneous light stimulus is implemented as follows. It is noted that the time course of the presented rate equation model always induces relaxation to a unique stationary state for a given parameter set such as reaction coefficients and $X^t_{i}$, independent of the initial concentrations of the signaling components. Here, the concentrations of signaling components of this stationary state correspond to those of the rod cells kept in darkness for a sufficiently long period of time. Thus, to consider the behaviors of this model in response to an instantaneous light stimulus, we set the state of the system at $t=0$ as follows. First, we make the system relax to the stationary state with long-term simulation. Here, $X_{R^*}$ relaxes to $=0$. Next, after inducing the relaxation, we change the concentration of $X_{R^*}$ from $0$ to $X_{R^*}^0$ ($X_{R}$ from $X_{R}^t$ to $X_{R}^t-X_{R^*}^0$), and set the time as $t=0$. Here, $L = X_{R^*}^0/X_{R}^t$ is considered to be proportional to the intensity of the light stimulus. Then, we define $L$ as the light intensity.

In experiments, the photoactivation of cells was estimated by monitoring the $Ca^{2+}$ flux through $Ca^{2+}$ channels, $\phi$, as $\displaystyle r = 1 - \frac{\phi}{\phi^{dark}}$, where $\phi^{dark}$ indicates the $Ca^{2+}$ flux in the darkness\cite{rod1,rod2,rod3, rod4,rod7}. In the present model, the $Ca^{2+}$ flux is proportional to $X_{Ch_o}(t)$. Thus, we measure
\begin{eqnarray}
r(t) = 1 - \frac{X_{Ch_o}(t)}{X_{CH_o}(0)}
\end{eqnarray}
to estimate the photoactivation of the system.

\subsection{Assumptions of the parameters for WT and mutant models}

The present rate equation model involves a large number of parameters such as the reaction coefficients and the total numbers of chemical components. Now, we assume the following parameter sets for the model to most accurately reproduce and explain several of the experimental results observed in mouse rod cells.

First, we set the total concentration of each signaling component $X^t_{i}$ according to the following assumptions for constructing both the WT model and mutant model of cells. As shown in recent experiments, the area fractions of rhodopsin are expected to be $\sim 0.27$ in WT mouse rod cells, in contrast to $\sim 0.135$ for mutant cells\cite{rod4,rod7}. It is noted that the area fration is restricted to be $\le 1$. In the previous subsection, we set the same restriction for the total concentration of signaling components on the disk membrane. Thus, we assume $X^t_{R} = 0.27$ for the WT model and $X^t_{R} = 0.135$ for the mutant model.

It has also been shown that the expression levels of genes encoding signaling proteins are almost identical between WT and mutant cells, except for rhodopsin\cite{rod4}. Thus, we assume that $X^t_{PDE}$, $X^t_{G}$, $X^t_{RGS9}$, and $X^t_{RK}$ are the same between the WT and mutant models, and their sum obeys $\le 0.4 - X^t_{R}$, since the volume fractions of molecules in normal cells are estimated to be $0.2 \sim 0.4$. In the following arguments, we mainly focus on the cases of $X^t_{G}=0.01$, $X^t_{PDE}=0.1$, $X^t_{RGS9} =0.0005$, and $X^t_{RK}=0.0005$, and the total concentration of $Rec$ in the cytoplasm and that of the channel on the cell membrane are assumed to be $X^t_{Rec}=0.05$ and $X^t_{Ch}=0.05$, respectively.

Next, we consider the reaction coefficients. The reaction coefficients are given as the combinations of the association and/or dissociation rates of molecules and their diffusion constants. However, not all of these values are precisely determined or reported in experiments. Thus, each reaction coefficient is set according to the following naturally assumed facts aiming for simplicity.

A1) In general, the diffusion of molecules on the lipid membrane is slower than that in the cytoplasm, suggesting that two-body reactions occur more frequently in the cytoplasm than on the membrane. Thus, the reaction coefficients for two-body reactions on the disk membrane are considered to be smaller than those in the cytoplasm. In this argument, for simplicity, we assume that the reaction coefficients on both the disk membrane and cytoplasm are respective constant values $k_{mem}$ and $k_{cyt}$, independent of signaling components, when the concentrations of the molecules are diluted.

A2) The diffusion of molecules in cells is restricted by their molecular crowding\cite{crowd2,crowd3,crowd4,crowd5,crowd6,crowd7,crowd8,crowd9,crowd10,crowd11,crowd12,crowd13,crowd14}. Thus, each reaction in a molecularly crowded environment is slower than that in an enviroment with diluted molecules. In this argument, we assume that the effective reaction rate of two-body reactions on the disk membrane and cytoplasm, $k'_{mem}$ and $k'_{cyt}$, obey $k_{mem} : k'_{mem} = 1:(1-\rho)$ and $k_{cyt} : k'_{cyt}= 1:(1-\rho_c)$, where $\rho_c$ refers to the total volume of molecules in the cytoplasm, and is assumed to be constant since the fluctuation of the molecular density in the cytoplasm seems to be small and relax immediately.

Then, the coefficients of the two-body reaction processes on the disk membrane are $A$, $B$, $C^{+}$, $D$, $E$, $S$, $T$ $= k'_{mem} = k_{mem}(1-\rho)$, and $H$, $J$, $P$, $Q$ $= k'_{cyt}$. Here, we regard the degradation of $cGMP$ by $(2G \cdot PDE)$ and the opening of the $Ca^{2+}$ channel by $cGMP$ (reaction (6)) as the representative cytoplasmic reactions, since the diffusion of $cGMP$ in the cytoplasm is an essential determinant of the frequency of these reactions. Moreover, the binding rate of $2Ca^{2+}\cdot Rec^c$ to the disk membrane is assumed to be similar to that of the reactions on the membrane, since this process is also influenced by the molecular concentration on the disk membrane.

A3) The diffusion of $Ca^{2+}$ in the cytoplasm is faster than any other reaction process. Here, we assume that the inflow and outflow rates of $Ca^{2+}$ are the same with $M = N = k_{ion}$, which is set as a large value.

A4) If any $cGMP$ molecules do not bind to the $Ca^{2+}$ channel, this channel is considered to close quickly. Thus, $K =k_c$ is set to be a large value.

A5) The spontaneous reactions are considered to be slower than catalytic reactions. For simplicity, we assume that the coefficients of spontaneous degradation of signaling components are small constant values of $C^{-}$, $F$, $I$, $R$ $=k_{d}$.

A6) The synthesis of molecules is also considered to be a relatively slower reaction. Thus, the production rate of $cGMP$, $G = k_{cG}$, is set to be a small value.

Based on these assumptions, we classify the parameters as follows: L) $k'_{cyt}$, $k_{ion}$, and $k_{c}$ have large values; S) $k_{d}$ and $k_{cG}$ have small values; and M) $k_{mem}$ has an intermediate value. In the following sections, we mainly focus on the results derived in the case of $k'_{cyt} = 1.6 \times 10^5$, $k_{mem} = 8.0 \times 10^4$, $k_{ion}= 1.6 \times 10^5$, $k_{c}= 1.6 \times 10^5$, $k_{d}= 4.0 \times 10^2$, and $k_{cG}= 4.0 \times 10^2$.

\section{Results and discussion}

First, we focus on the time course of $r(t)$ for both the WT and mutant models of mouse rod cells for an appropriately large $L$ (Fig. 2(a,b)). The results showed that the relaxation of the mutant model was faster than that of the WT model for each $L$, which is consistent to experimental observations\cite{rod4,rod7}. Notably, the photoactivated state with large $r(t)$ was prolonged with the appearance of the plateau of $r(t)$ for large $L$ values, a result that has not been obtained in recent theoretical models of the mouse rod cell\cite{rod6,rod7,rod8,rod9,rod10,rod11,rod12,rod13,rod14} but has been observed in experiments\cite{rod1,rod2,rod3,rod4,rod7}.

\begin{figure}
\begin{center}
\includegraphics[width=12.0cm]{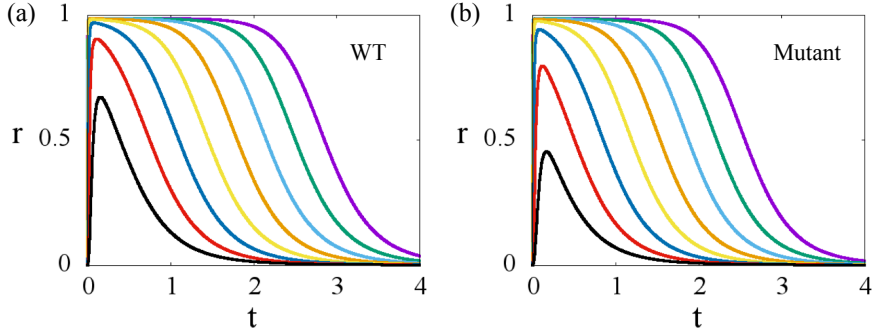}
\end{center}
\caption{{\bf Time courses of photoactivation of WT and mutant model of mouse rod cell: }  Time courses of the photoactivation, $r(t)$, of the (a) WT and (b) mutant models of the mouse rod cell. The difference of the colors of curves indicate variations in $L$ where purple, green, sky blue, orange, yellow, blue, red, and black curves indicates the cases of $L = 1, 0.75^4, 0.75^8, ... , 0.75^{28}$.}
\end{figure}

Second, the lifespan of the photoactivation of cells, $\tau$, was measured, which is defined as the time interval from the point at which $r(t)$ reaches its maximum $r_{max}$ to the point at which $r(t) = 0.8r_{max}$, based on a recent experiment\cite{rod4} (Fig. 3(a)). For the cases in which $L$ is larger (than $\sim 2.0 \times 10^{-4}$), $\tau$ for the WT model was always longer than that for the mutant model for the same $L$, as observed in the experiment\cite{rod4}. Moreover, in both models, the slope of $\tau$ increased with an increase in $L$, and $tau$ showed a logarithmic increase when $L$ was so large that $r(t)$ reached a plateau. Such phenomena were also observed in the experiment\cite{rod4}. In addition, both the activation and relaxation of the mutant model were faster than those of the WT model in the case of the same $X_{R^*}^0$ ($L=L_w$ for the WT model and $L = 2L_w$ for the mutant model) when $L$ is larger than $\sim 2.0 \times 10^{-4}$ but $X_{R^*}^0$ is small (Fig. 3(d)), as observed in experiments with weak light intensity\cite{rod4,rod7}.

\begin{figure}
\begin{center}
\includegraphics[width=12.0cm]{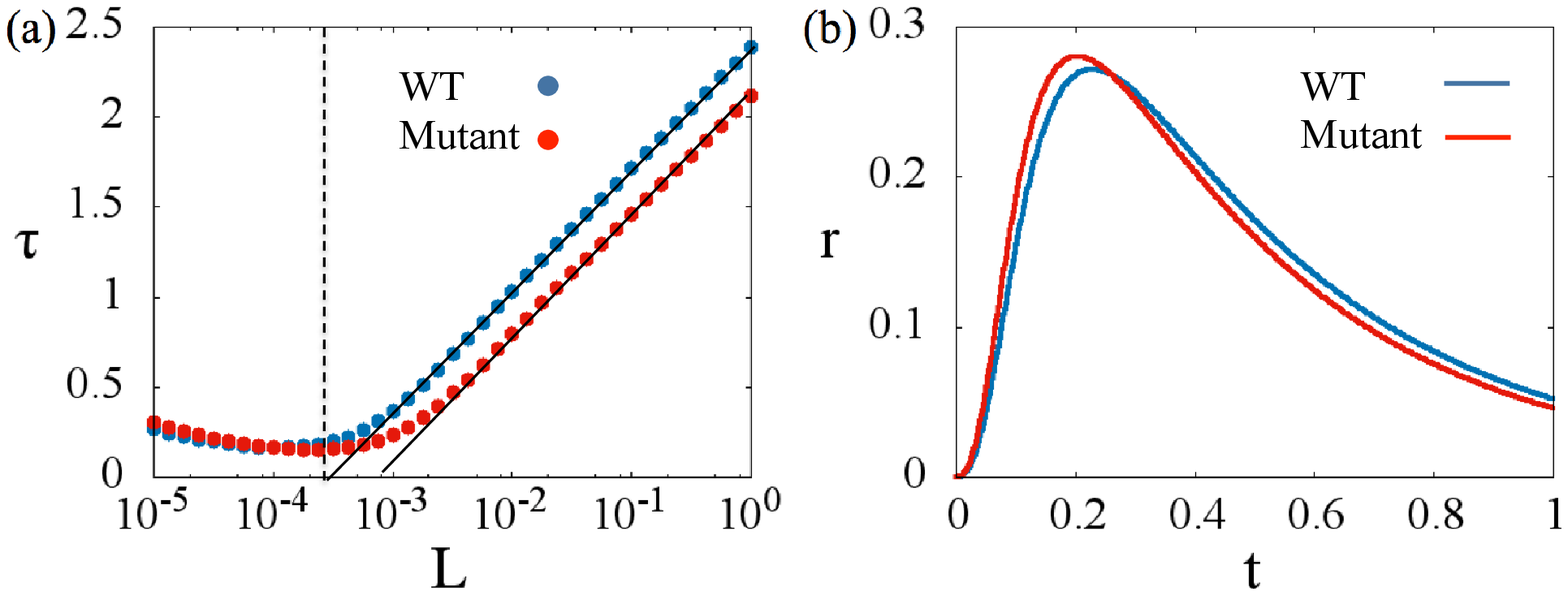}
\end{center}
\caption{{\bf Light intensity dependency of the lifespan of photoactivation time, and time courses of the photoactivation in the WT and mutant models given weak light stimuli: } (a) Lifespan of the photoactivation $\tau$ as a function of light intensity $L$ in the WT (blue) and mutant (red) models, and (b) time courses of $r(t)$ of the WT (blue curve) and mutant (red curve) models for the same small value of $X_{R^*}^0$ ($= 2.41 \times 10^{-4}$).}
\end{figure}

The present model also demonstrated a decrease in $\tau$ for smaller $L$ values (less than $\sim 2.0 \times 10^{-4}$) with an increase in $L$ (Fig. 3(a)). Such behavior has not been reported in any experiments. We speculate that this was due to acceleration in the relaxation of the activation of each rod cell with increasing light intensity within the context of very weak light stimuli. However, this conjecture should be confirmed in future experiments to evaluate the validity of the present model.

\begin{figure}
\begin{center}
\includegraphics[width=12.0cm]{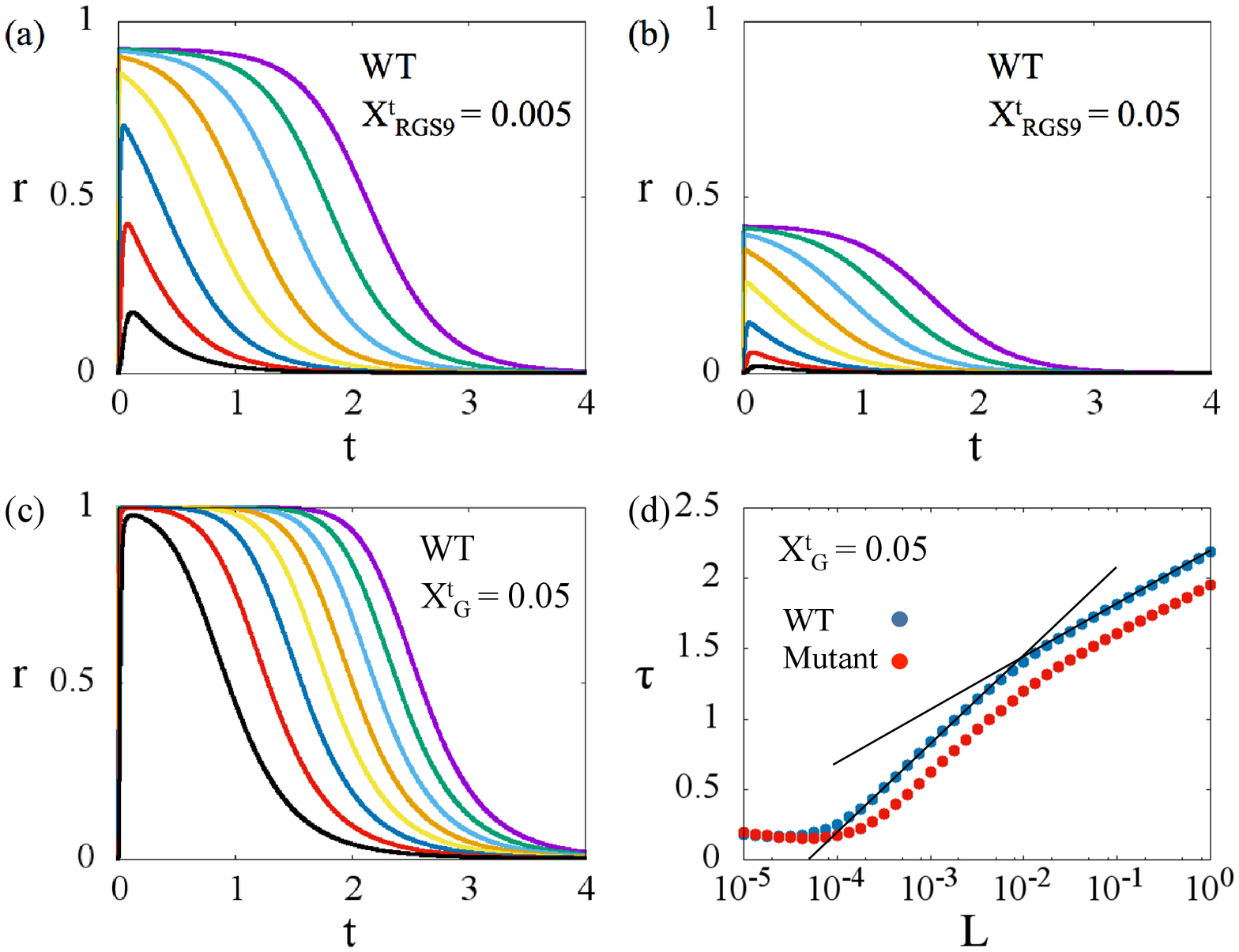}
\end{center}
\caption{{\bf Time courses of photoactivation and the light intensity dependency of the lifespan of  photoactivation for the cases of different parameter sets from the ideal case: } The time courses of $r(t)$ of the WT model for the cases of $X^t_{RGS9}$ increase to (a) $0.005$ and (b)$0.05$. The colors of the curves are the same as those described for Fig. 2. (c) Time courses of $r(t)$ of the WT model and (d) the lifespan of the photoactivation $\tau$ as a function of light intensity $L$ in the WT (blue) and mutant (red) models for the case of $X^t_{G}$ increases to $0.05$.}
\end{figure}

Now, we consider the mechanism of the phenomena obtained in the present model. In particular, we focus on the accelerations of the activation and relaxation due to the mutations and the appearances of the plateau of $r(t)$ for large $L$.

The reason for the accelerations in the activation and relaxation in the mutant cell with a reduced rhodopsin density can be easily understood as follows. The limiting steps of the present signaling processes are the two-body reactions on the disk membrane, becauase of the slow diffusion and molecular crowding effects on the membrane. In the mutant model, the concentraton of rhodopsin is half that of the WT model. Thus, the reaction coefficients for the limiting reaction steps of the mutant model are larger than those of the WT model. By this simple fact, the photoactivation and relaxation of mutant cells are more quickly achieved than those of WT cells.

Next, we consider the mechanism of the slow relaxation of $r(t)$ with the appearance of the plateau for large $L$ values. The dominant contribution of the appearance of the plateau is considered to result from the deficiency of $RGS9$; in particular, its concentration is substantially reduced compared to $G$ and $PDE$, which slows down the degradation of $(2G^* \cdot PDE)$ to $2G + PDE$. If $L$ is small, the concentration of $G^*$ is not so large, so that the concentration of $(2G^* \cdot PDE)$ is similar or less than that of $RGS9$. In this case, $(2G^* \cdot PDE)$ is rapidly transformed to $(2G^* \cdot PDE \cdot RGS9)$ and the degradation of $cGMP$ is immediately suppressed. On the other hand, if $L$ becomes large, the concentration of $G^*$ becomes so large that the concentration of $(2G^* \cdot PDE)$ becomes larger than that of $RGS9$. Consequently, most of the $RGS9$ becomes involved in $(2G^* \cdot PDE \cdot RGS9)$, and $cGMP$ continues to be degraded by the remaining $(2G^* \cdot PDE)$. Thus, the transformation from $(2G^* \cdot PDE)$ to $(2G^* \cdot PDE \cdot RGS9)$ is hindered by the conflict of $RGS9$, which induces the observed plateau of $r(t)$ for large $L$

These arguments were confirmed by simulations of the model using larger values of $X^t_{RGS9}$ than $0.0005$ (Fig. 4(a, b)) where the length of the plateau, the range of $L$ with which the plateau appears, was changed. Moreover, the maximum $r(t)$ was reduced with an increase in $RGS9$. Thus, the deficiency of $RGS9$ has a dominant contribution to the amplification and reduction in the rate of relaxation of photoactivation of rod cells.

Finally, we show the typical behaviors obtained by the present model when the parameters are changed from the previously considered ideal cases. If $X^t_{G}$ ($X^t_{PDE}$) becomes so large (small) to approach $X^t_{PDE}$ ($X^t_{G}$) or $X^t_{RK}$ becomes larger, a slow increase in $\tau$ with an increase in $L$ appeared again for larger $L$ values (Fig. 4(c,d)). Similar results were obtained when some reaction coefficients were randomly changed. However, these behaviors have not yet been reported experimentally.

\section{Conclusion}
In this study, we investigated the mouse rod phototransduction process through the development of rate equation models consisting of the chemical components that contribute to the signaling processes into and around the membranous disks in the outer segments of rod cells. By considering the effects of the molecular crowding of rhodopsin on the disk membrane, we could explain the mechanism observed both experimentally and theoretically by which photoactivation and relaxation in the WT mouse rod cell are slower than those of the mutant cell containing half the amount of rhodopsin. We also clarified that the deficiency of $RGS9$ compared to the other molecules on the disk membrane is primarily responsible for inducing prolongation of the photoactivated state of rod cells in response to strong light stimuli; these observations and explanations have not previously been obtained in recent theoretical models.

The present model includes a large number of parameters, and only a portion of these parameters are experimentally derived. Thus, in the presented arguments, we have adopted simple but reasonably appropriate assumptions for estimating the unknown parameters. The accumulation of further experimental data for several of the relevant parameters in the future would allow us to address quantitative considerations and derive several predictions for the realistic behaviors of mouse rod cells. Moreover, the reduction and mathematical analysis of a few effective valuables and their rate equations from the presented model should be conducted in the future in order to unveil more detail properties and mechanisms of these signaling processes.

\section*{Acknowledgment}
The author is grateful to M. Fujii, T. Yamashita, and M. Yanagawa for fruitful discussions. This research is partially supported by the Platform Project for Supporting in Drug Discovery and Life Science Research Platform for Dynamic Approaches to Living System from the Ministry of Education, Culture, Sports, Science (MEXT), the Japan Agency for Medical Research and development (AMED), and the Grant-in-Aid for Scientific Research on Innovative Areas ``Spying minority in biological phenomena (No.3306) (24115515)'' and ``Initiative for High-Dimensional Data-Driven Science through Deepening of Sparse Modeling (No.4503)(26120525)'' of MEXT of Japan.

\end{document}